\documentstyle[11pt,aasms4]{article}

\begin{document}

\title{SAX J1810.8-2609: A New Hard X-ray Bursting Transient 
      }
\author{L. Natalucci\altaffilmark{1}, A. Bazzano, M. Cocchi, and P. Ubertini}
          \affil{Istituto di Astrofisica Spaziale {\em(IAS/CNR)},
           via Fosso del Cavaliere, 00133 Roma, Italy}
\author{J. Heise, E. Kuulkers\altaffilmark{2}, J.J.M. in 't Zand and
           M.J.S. Smith\altaffilmark{3}} 
          \affil{Space Research Organization Netherlands {\em (SRON)},
           Sorbonnelaan 2, 3584 CA Utrecht, The Netherlands}

\altaffiltext{1}{e-mail address: lorenzo@ias.rm.cnr.it}
\altaffiltext{2}{Also: Astronomical Institute, Utrecht University, P.O.\ Box 80000,
3507 TA, Utrecht, The Netherlands}
\altaffiltext{3}{Also: BeppoSAX Science Data Centre, Nuova Telespazio, via
Corcolle 19, 00131 Roma, Italy}

\begin{abstract}
 
The transient X-ray source SAX~J1810.8-2609 was discovered on
1998, March 10 with the Wide Field Cameras 
on board the {\em BeppoSAX} satellite, while observing the 
Galactic Bulge in the 2-28 keV energy range. On March 11,  
a strong type-I X-ray burst was detected with evidence of photospheric
radius expansion.
A follow-up target of opportunity observation with the Narrow-Field
Instruments (NFI) was performed on March 11 and 12, for a total 
elapsed time of $8.51\times10^{4}$s. The wide band spectral data
(0.1-200 keV) obtained with the NFI show a remarkable hard X-ray spectrum 
detected up to $\sim$~200~keV, which can be described by a power law with   
photon spectral index $\Gamma$=1.96$\pm$0.04, plus a soft component which is 
compatible with blackbody radiation of temperature kT~$\sim$~0.5~keV.

The detection of the type-I X-ray burst is a strong indication that the compact
object is a neutron star in a low mass X-ray binary system. Assuming standard
burst parameters and attributing the photospheric radius expansion to 
near Eddington luminosity, we estimate a distance of $\sim$~5~kpc. 
The inferred 2-10~keV X-ray luminosity is   
$\sim$~$9\times10^{35}$ erg~s$^{-1}$ at the time of the discovery.

\end{abstract}

\keywords{binaries: close, individual ({\em SAX~J1810.8-2609})
          --- X-rays: bursts}

\section{INTRODUCTION}

During a long term 2-28 keV monitoring campaign of the Galactic Bulge region
with the Wide Field Cameras (WFC) on board the {\em BeppoSAX} satellite, the
new X-ray transient SAX~J1810.8-2609 was discovered on 1998, March 10
(\cite{Ube8a}). The source showed a  
weak emission ($\approx$15~mCrab) corresponding to an X-ray flux of
$3.1\times10^{-10}$~erg~cm$^{-2}$~s$^{-1}$ in the 2-10 keV range and was 
positioned in quasi real-time with
the quick-look analysis (QLA) tools at $\alpha=18^{h}10^{m}46^{s}$ and 
$\delta=-26\arcdeg 09\arcmin .1$
(equinox 2000.0) with an error radius of $3\arcmin$.
During the on-going  monitoring, on March 11 a strong  
type-I X-ray burst was observed with a peak intensity of $\sim1.9$~Crab, from a
sky position consistent with that of the  
persistent emission (\cite{Coc99}, \cite{Ube8a}). Two days after the source
was discovered a follow-up observation was performed with the {\em
BeppoSAX} Narrow Field Instruments
(NFI) showing that the 2-10~keV intensity had declined to
$\approx7.5$~mCrab (\cite{Ube8b}).  
On 1998, March 24 the {\em ROSAT}~High Resolution Imager (HRI) observed the
error box of SAX~J1810.8-2609 for 1153~s (\cite{Gre98}). A low energy source,
named  RX~J1810.7-2609, was detected at 
a position consistent with the WFC error box but not with the one 
obtained by the QLA of the NFI observation.
(\cite{Ube8b}, see however further details in Sect.2.1).  
The 0.1-2.4 keV flux of RX~J1810.7-2609 was $\sim1.5$~mCrab and {\em
ROSAT} did not detect the source in previous observations of the same sky
region on 1993 September 10 (0.1-2.4 keV), with a 3$\sigma$ upper limit of
$\sim 0.08$   mCrab and in 1990 during the All-Sky Survey, thus confirming the
transient nature  of the source. Very recently, Greiner et al. (1999) have
reported details of the {\em ROSAT}~HRI target of opportunity (TOO), and of
optical to infrared follow-up observations of the 20$\arcsec$ error box of the
{\em ROSAT}~HRI source. They tentatively suggested  as counterpart of
RX~J1810.7-2609 a variable object showing $R=19.5\pm0.5$ on March 13 and
R~$>$~21.5 on August 27.  The {\em ROSAT}~HRI observation showed an unabsorbed
flux of $\sim$~$1.1\times10^{-10}$~erg~cm$^{-2}$~s$^{-1}$. If one assumes a 
Crab-like spectrum, this extrapolates  
to $\sim$~$3.5 \times10^{-11}$~erg~cm$^{-2}$~s$^{-1}$ in the 2-10 keV range, 
which is a factor of 4 lower than the {\em BeppoSAX}~NFI detection
(\cite{Ube8b}, \cite{Gre99}). The variability during the 
{\em ROSAT}~HRI observation was less than a factor of 3 in the 0.1-2.4 keV and
no evidence of coherent or Quasi-Periodic Oscillations (QPO) in the range from
2 to 200~s was found, with a 3~$\sigma$  upper limit on the pulsed fraction of
$<$~40~\% (Greiner et al., 1999).

We here report on a detailed analysis of the WFC and NFI observations of 
SAX~J1810.8-2609, and discuss the nature of the compact source in this X-ray
transient.

\section{OBSERVATIONS AND DATA ANALYSIS}

The WFC (\cite{Jag97}) on board {\em BeppoSAX} are
designed for performing spatially resolved simultaneous measurements of X-ray
sources in crowded fields enabling studies of spectral variability  at high
time resolution. The mCrab sensitivity in 2-28~keV over a
$40\times40$~square degrees field of view (FOV) and the near-to-continuous
operation over a period of years offer the unique
opportunity to measure continuum emission as well as bursting behaviour from
many new X-ray transients and already known (weak) transient and persistent
sources.  For this reason the Galactic Bulge is being monitored over 1 to 2
months during  each of the  visibility periods since
the beginning of the {\em BeppoSAX} operational life in July 1996. During those
observations, that combine to a total of $\sim3$~Ms net exposure time up to 
November 1999, more than 900 X-ray bursts
and at least 45 sources have been detected
(\cite{Coc98}; \cite{Hei99}; \cite{Ube99}).  The data of the two cameras are
systematically searched for bursts or flares by analysing  the time  profiles
in the 2-10~keV energy range with a 1~s time resolution.

Follow-up observations with the more sensitive, broad band Narrow Field
Instruments are often performed each time a new transient source
is detected in the WFC field of view. 
The {\em BeppoSAX}~NFI comprise an assembly of four imaging instruments: one 
low energy and three medium energy concentrator spectrometers, 
named LECS and MECS, with a 37 and 56 arcmin circular FOV and  
energy ranges 0.1-10 keV and 1.8-10 keV, respectively 
(Parmar et al., 1997 and Boella et al. 1997). The other two non-imaging
co-aligned detectors are the High Pressure Gas Scintillation Proportional Counter
(HPGSPC), operative in the range 4-120 keV (Manzo et al., 1997) 
and the Phoswich Detector System (PDS), operative in the range 
15-200 keV (Frontera et al., 1997). 
On 1998, March 12.19 UT a {\em BeppoSAX}~follow-up observation was performed
with the NFI on the WFC error box of the newly discovered source 
(Ubertini et al., 1998b). The total observation lasted 85.1 ks corresponding 
to a net exposure time of 14.4~ks for LECS, 26.8~ks for MECS, 20.0~ks for
HPGSPC and 30.4~ks for PDS.  SAX~J1810.8-2609 was 
strongly detected in all instruments, including a high energy tail extending up 
to $\sim$~200~keV and was the only source present in the 
LECS and MECS images, at an updated position consistent with the WFC error
box (see Sect. 2.1).
Extraction radii of $8\arcmin$ an $4\arcmin$ have been used for source photons
for the LECS and MECS images
respectively, encircling $\sim 95$~\% of the power of the
concentrators point spread function. These data have been
used for spectral analysis and light curves production.
All spectra have been rebinned, oversampling the detector spectral resolution,
to have at least 20 counts per channel.
The bandpasses
for spectral analysis were limited to 0.3-3.0~keV for the LECS, 1.6-10.5~keV
for the MECS, 4-25~keV for HPGSPC and 15-200~keV for the PDS to take advantage
of accurate detectors calibration. The standard procedure to leave free the
relative normalization parameters of the different instruments within a narrow
band, was  applied, to accommodate cross-calibration uncertainties. 

\subsection{The source position}
On March 11.06633 UT a strong burst was observed from SAX~J1810.8-2609; this is the 
only X-ray burst ever observed from the source in all the WFC
data since 1996, which amounts to a total net exposure time of $\sim3$~Ms.
We have improved the source position in the WFC  
with respect to the one previously reported (\cite{Ube8a}) to 
$\alpha=18^{h}10^{m}45.6^{s}$ and $\delta=26\arcdeg 08\arcmin 48.5\arcsec$
(1.1~$\rm arcmin$ error radius), by using the burst data which has a much
higher statistical quality than that of the non-burst data (see Table 1).
This confirms the association with the {\em ROSAT}~HRI source 
RX~J1810.7-2609.
We note that the original inconsistency between the {\em BeppoSAX}~NFI and 
{\em ROSAT}~HRI (\cite{Gre98}, \cite{Ube8b}) was 
due to an error in the aspect solution of {\em BeppoSAX} which resulted from
an unusual attitude configuration.
We have therefore refined the position of the source taking into account 
a new calibration (L.~Piro, L.A.~Antonelli, private communication). This
results in $\alpha=18^{h}10^{m}45.5^{s}$  
and $\delta=-26\arcdeg 08\arcmin 14\arcsec$
(equinox 2000.0) with a conservative error radius of $1.5\arcmin$, and
is now consistent with that determined by the {\em ROSAT}~HRI. 
The various error circles are shown in Figure 1. 


\subsection{The single X-ray burst}

A single, strong burst was detected from SAX~J1810.8-2609 on 1998 March
11.06634. The event lasted 47~s with an e-folding time of $12.5\pm0.7$~s and
showed a peak intensity of $1.9\pm0.2$~Crab in the 2-28~keV band 
(see also \cite{Coc99}). The time
profiles in two energy bands are shown in Figure 2: a clear double-peaked
structure is present at high  energy (10-28 keV) suggesting photospheric
radius expansion (\cite{Lew95}). The spectrum of the burst obtained
integrating data over the whole  burst duration  is well represented by a
blackbody emission with temperature kT~$\simeq$~2 keV. In order to study the
time resolved spectra we have integrated the burst data   in time intervals as
shown in the lower panel of Fig.2, more or less corresponding to the peak
structures observed in the high energy profile. 
Under given assumptions (\cite {Lew93}) the effective temperature
${T}_{eff}$ and the bolometric flux of a burst can determine the ratio between
the blackbody radius ${R}_{bb}$ (that is, the radius of the emitting sphere)
and the distance d of the neutron star. Assuming $d=10$~kpc and the observed 
colour temperatures as ${T}_{eff}$, and not correcting for gravitational
redshift the data 
are consistent with a radius expansion of a factor of $\sim2$ during the first
$\sim10$~s of the event. The average blackbody radius, excluding the 
radius expansion part is $\sim12$~km (see Table 2) at 10 kpc.
Also evident is the 
typical spectral softening  due to the cooling of the photosphere after the
contraction of the emitting region. These results clearly indicate that the
burst is of type-I, i.e. it is identified as a thermonuclear flash on a neutron
star (NS). The total bolometric fluence of the burst, estimated by spectral analysis
is $(1.45\pm0.06)\times10^{-6}$~erg~cm$^{-2}$.


The observation of the near-Eddington profile is a
clue to estimate the source distance. In fact, for a 1.4~${M}_{\odot}$~NS
and a corresponding Eddington bolometric luminosity of $2\times10^{38}$~erg/s
we obtain d=($4.9\pm0.3$)~kpc, assuming standard burst parameters (here the
error is purely statistical). For
this distance the total burst emitted energy is $\sim4\times10^{39}$~erg and
the observed blackbody radius scales to a value of $\sim6$~km. This value of
radius could be underestimated, 
due to the uncertainties in the relationship between colour and effective 
temperature. If,
as suggested by \cite{Ebi87} the colour temperature exceeds ${T}_{eff}$ by a
factor $\approx$1.5, then the neutron star radius should be at least two
times the measured blackbody radius. These values therefore support 
a neutron star nature of the compact object.


     
\subsection{The wide band persistent emission}

The light curve of SAXJ1810.8-2609 measured with the {\em BeppoSAX}~NFI is
shown in Figure 3, in different energy ranges. There is a slight decrease
of the flux in the lower energy range (E$<$10 keV) in the first $\sim60$~ks
of the observations, while there is no clear evidence for a decline  
in the final part of the observation. This picture is consistent with the 
overall flux trend of this source, and with the derived 
e-folding time of $\sim~7.5$ days that is estimated from the WFC, 
{\em ROSAT}~HRI and NFI observations.


The count rate spectrum shows substantial emission
at high energy.
In fact, the unfolded spectrum in the energy range 15-200~keV can be fitted 
by a single power law of spectral index
$\Gamma=2.02\pm0.07$ ($\chi^{2}_{r}$=0.76 over 15 d.o.f),
and a flux in this range of $2.2\times10^{-10}$~erg~cm$^{-2}$~s$^{-1}$.
The broad band spectral data, fitted by a single absorbed power law
results in a photon spectral index of $2.22\pm0.02$, with a reduced chi-square
$\chi^{2}_{r}$=1.35 over 165 degrees of freedom
and an average flux of $4.2\times10^{-10}$~erg~cm$^{-2}$~s$^{-1}$
in the 0.1-200~keV band. It is clear that
the absorbed power law model is not satisfactory when applied to the broad band
emission.

The fit is significantly improved
by using a thermal comptonization spectrum ({\it comptt} in XSPEC v.10)
instead of the simple power law, resulting in $\chi^{2}_{r}$=1.12 for 163 d.o.f.
(see Table 3) which corresponds to a null hypothesis probability of 0.147.
In this model, the hard X-ray tail is produced by the upscattering of 
soft seed photons by a hot, optically thin electron plasma (\cite{Sun80}). 
The seed photons temperature for this fit is ($0.36\pm0.02$)~keV. 
The hard X-ray data, however, cannot constrain the  
parameters of the Compton emission region (temperature and optical depth)
due to the very high energy cutoff which is above $\sim150$~keV.

The addition of a soft thermal component  
improves both the power law and comptonization fits. The soft component can 
be modelled satisfactorily with blackbody or multicolor disk (MCD) blackbody 
emission 
(\cite{Mit84}). Using single temperature blackbody, the fits for power law and  
comptonization are both compatible with a temperature value
kT$\approx$~0.5~keV (see Table 3 for details), giving a $\chi^{2}_{r}$ of 0.97 and 
0.99 respectively. 
The power law photon spectral index is $\Gamma$=$1.96\pm0.04$ and the temperature 
of the soft comptonized emission is ($0.6\pm0.4$)~keV. 
The estimated blackbody flux is between $\sim2.5$ and 
$\sim$~4~$\times$10$^{-11}$~erg~cm$^{-2}$~s$^{-1}$. At the quoted
source distance of 4.9~kpc this indicates an emission radius between 
$\sim10$ and $\sim40$~km. Using a MCD model to describe the additional soft 
component, the thermal emission is characterized by ${kT}_{in}$=$0.6\pm0.1$~keV   
(temperature at the inner disk radius ${R}_{in}$). For this model, the best fit 
gives $\chi^{2}_{r}$=0.99 for 161 d.o.f. The values  
for ${R}_{in}$~$\sqrt{\cos{\theta}}$ may range from $\sim1.5$ to $\sim10$~km
(here $\theta$ is the disk viewing angle). Hence, if this soft component
originates from an optically thick region of the accretion disk, this should be 
expected to be not too far from the NS, unless the disk is seen at very large
inclination.  

The broadband source spectrum unfolded by the four instruments response is 
shown in Fig.4 along with the model spectrum obtained for the blackbody 
component plus thermal comptonization best fit. 
We note that the value of ${N}_{H}$~$\simeq$ 3.5$\times$10$^{21}$~cm$^{-2}$
obtained for the fits which include comptonization match very well the
current estimate of the Galactic column density of
$\approx3.7\times10^{21}$~cm$^{-2}$ for this region (\cite{Dic90}). 


\section{DISCUSSION}

The deep and timely investigations carried out by means of repeated 
{\em BeppoSAX} observations
of SAX~J1810.8-2609 are consistent with a transient type-I X-ray
bursting source, most likely a low mass X-ray binary (LMXB) containing a weakly
magnetized NS. This source is a weak transient, as supported by the fact
that it was never detected in more than three years of {\em
BeppoSAX} monitoring of the Galactic Bulge region (apart from these discovery
and follow-up observations)
and also never seen by the {\em RXTE}~All Sky Monitor (ASM), even during the
March 1998 outburst.  The ASM non-detection implies an upper limit on the
2-10~keV flux of $\sim7\times10^{-10}$~erg~cm$^{-2}$~s$^{-1}$. It is 
noteworthy, that a similar weak transient behaviour has also been observed in
a number of recently discovered bursters, detected during dim
X-ray outburst episodes with maximum intensities well below 100 mCrab and
lasting $\sim1$ to a few weeks (see e.g. \cite{Hei99}). 

The estimated value of distance of $\sim5$~kpc, that we obtained from the 
observation of radius expansion during the burst (\cite{Lew93}; 
\cite{Lew95}) places SAX~J1810.8-2609 at our side of the Galactic Bulge 
(see e.g. \cite{Chr97}). This is consistent with the tentative detection
of the optical counterpart (\cite{Gre99}). We note that the presence of the 
neutron star in the system is also supported by the relatively small 
blackbody radius of $\sim6$~km, calculated for the derived distance.     

The detection of a single X-ray burst during our monitoring observations
is consistent with the observed combination of burst fluence and average
persistent bolometric emission, which is 
$\sim5\times10^{-10}$~erg~cm$^{-2}$~s$^{-1}$, i.e. $\sim0.01$~${L}_{edd}$ at
5~kpc. In fact, taking into account the total energy release of the burst 
and assuming that steady nuclear burning is negligible, we
can estimate a typical value of $\sim5$~days for the mean burst interval,
which corresponds to the expected $\alpha$ parameter for helium burning (i.e.,
$\alpha$~$\geq100$, see \cite{Lew93}) and is comparable with the  
e-folding decay time of 7.5 days estimated for the persistent emission.
Conversely, if a significant part of the nuclear fuel is burnt steadily the
quoted value should be considered as a lower limit.


The broad band spectrum of SAX~J1810.8-2609 shows a   
high energy power law tail, which is   
remarkably hard ($\Gamma=2.06\pm0.11$ in the 15-200~keV band) and with no 
cutoff. There is
also an indication for a soft blackbody component with temperature 
$kT\approx0.5$~keV and total  
flux ${F}_{bb}$~$\sim$~3~$\times$10$^{-11}$~erg~cm$^{-2}$~s$^{-1}$.
The ratio of the soft component luminosity to the total X-ray (0.1-10 keV)
luminosity is estimated to be in the range $\sim$~10-15\%. This is consistent 
with upper limits obtained for X-ray bursters observed by ASCA 
in the low state (see e.g., \cite{Rev99}) and also with the detection of
similar  soft components in the spectra of 4U~0614+091 (\cite{Pir99}),
1E~1724-3045 and   SLX~1735-269 (\cite{Bar99}), which were all observed in a
hard state.  A recent analysis of {\em ROSAT} spectra of LMXB (Schulz 1999)
also shows that a soft component is present in several low luminosity (mainly,
Atoll type) X-ray bursters. 

The luminosities in the soft and hard X-ray bands match quite well the
observed  correlation pattern found for neutron star binaries in the low state 
(\cite{Bar99}), with values of $\sim$~$7.5\times10^{35}$ erg~s$^{-1}$
in the 1-20~keV band, and $\sim$~$5.0\times10^{35}$ erg~s$^{-1}$ in the 
20-200~keV band. Nevertheless, the
absence of cutoff below $\sim$200~keV is particularly outstanding, as in
most cases X-ray bursters with hard tail spectra do have this feature which is
suggestive of comptonization with plasma temperatures below $\sim$~50~keV
(\cite{Gua98}, \cite{Zan99}). The presence of such a cutoff was suggested as
a possible criterion to distinguish NS from black hole (BH) spectra, the latter
being characterized by much higher electron temperatures (\cite{Tav97}).
The case of SAX~J1810.8-2609 is not compatible 
with this kind of interpretation.  Very
recently, an analysis of {\em BeppoSAX} observations of the Atoll X-ray burster 
4U~0614+091 has revealed a similar behaviour, i.e. a high energy power law
tail with no visible cutoff (\cite{Pir99}).  
Whether the spectrum of SAX~J1810.8-2609 could have
a cut-off just above 200 keV (that is our observational upper energy
limit) is difficult to say. Our broad band spectral analysis
shows that only a Comptonization fit is compatible with a low  energy
absorption matching the value of Galactic column density. We conclude that,
even if the data are not able to constrain the parameters of 
the scattering region we still have good indication that comptonization is 
the mechanism that produces the hard X-ray tail.

\acknowledgments

We thank Team Members of the BeppoSax Science Operation Centre and Science Data 
Centre for continuos support  
and timely actions for 
quasi 
real-time detection 
of new transient and bursting sources and the follow-up TOO  
observations. The {\em BeppoSAX} satellite is a joint Italian and Dutch programme. 
LN is grateful to D. Barret for useful discussion and specially thanks
A. Santangelo and A. Segreto for suggestions and help on HPGSPC data analysis.


\clearpage

\begin{center}
\begin{deluxetable}{crrrrrrrrrr}
\footnotesize
\tablecolumns{8}
\tablewidth{0pt}
\tablecaption{All available SAXJ1810.8-2609 observations, positions and associated
 error boxes.}
\label{tbl1}
\tablehead{
\colhead{Detector} & \colhead{Energy}  &
\colhead{Observation}  & \colhead{R.A.} &
\colhead{Decl.}  & \colhead{Error circle}  & \colhead{References}\\ 
\colhead{}  & \colhead{(keV)}  & \colhead{1998}&
\colhead{2000.0}  & \colhead{2000.0}  & \colhead{(99\% conf.)} & \colhead{}
}
\startdata
{\em BeppoSAX}~WFC & 2-28 & March 11\tablenotemark{a} & 
$18^{h}10^{m}45.6^{s}$ & 
$-26\arcdeg 08\arcmin 48\farcs5$ &
$1.1 \arcmin$ &
This work,\\ 
& & & & & & Cocchi et al.1999 \\
{\em BeppoSAX}~WFC & 2-28 & March 10-12\tablenotemark{b,c} &  
$18^{h}10^{m}46.9^{s}$& 
$-26\arcdeg 09\arcmin 24\farcs0$ & 
$2.0\arcmin$ &
This work\\
{\em BeppoSAX}~NFI & 0.1-200 & March 12-13\tablenotemark{b} & 
$18^{h}10^{m}45.5^{s}$ & $-26\arcdeg 08\arcmin 14\farcs0$ & 
$1.5\arcmin$ &
This work\\
{\em ROSAT}~HRI & 0.1 - 2.4 & March 24\tablenotemark{b} & 
$18^{h}10^{m}44.5^{s}$ & $-26\arcdeg 09\arcmin 0\farcs1$ & 
$20 \arcsec$ &  
Greiner et al.,1999\\
\enddata
\tablenotetext{a}{Burst emission}
\tablenotetext{b}{Persistent emission}
\tablenotetext{c}{Source discovery}
\end{deluxetable}
\end{center}



\begin{center}
\begin{deluxetable}{crrrrrrrrrrr}
\label{tbl2} 
\footnotesize
\tablecolumns{5}
\tablecaption{Burst fit parameters}
\tablewidth{0pt}
\tablehead{
\colhead{Data Period\tablenotemark{a}} & \colhead{Int. Time}   
  & \colhead{${kT}_{bb}$(keV)}  
  & \colhead{${R}_{bb}$(km)\tablenotemark{b}} 
  &  \colhead{$\chi^{2}_{r}$(26 d.o.f.)}
}
\startdata
Whole Burst & 48 s & $1.98\pm0.04$ & $12.3\pm0.6$  & 1.89  \\
First Peak  & 4 s  & $2.55\pm0.11$ & $9.5\pm1.1$   & 1.46  \\
First Tail  & 6 s  & $1.83\pm0.06$ & $20.1\pm1.7$  & 1.29  \\
Second Peak & 6 s  & $2.71\pm0.09$ & $9.7\pm0.8$   & 0.98  \\
Second Tail & 7 s  & $1.94\pm0.08$ & $14.6\pm1.6$  & 1.09  \\
Third Tail  & 25 s & $1.53\pm0.07$ & $13.2\pm1.7$  & 1.57  \\
\enddata
\tablenotetext{a}{refer also to lower panel of Fig.2}
\tablenotetext{b}{estimated for a distance of 10 kpc (Cocchi et al. 1999)}
\end{deluxetable} 
\end{center}


\clearpage
\begin{center}
\begin{deluxetable}{crrrrrrrrrrr}
\scriptsize
\tablecaption{Parameter values of spectral models fitted to the persistent 
emission}
\label{tbl3} 
\tablewidth{0pt}
\tablecolumns{11}
\tablehead{
         \colhead{Model} 
       & \colhead{Range\tablenotemark{a}}
       & \colhead{$\rm N_{H}$\tablenotemark{b}}
       & \colhead{$\rm C_{pl}$\tablenotemark{c}}
       & \colhead{$\Gamma$}
       & \colhead{${kT}_{0}$\tablenotemark{d}}
       & \colhead{${kT}_{e}$\tablenotemark{e}}
       & \colhead{$\tau$\tablenotemark{f}}
       & \colhead{${kT}_{bb}$\tablenotemark{g}}
       & \colhead{Flux\tablenotemark{h}}
       & \colhead{$\chi^2_\nu$\tablenotemark{i}} 
}   
\startdata
PL                   & 0.1-200 & $7.5\pm0.2$ &
  $8.5\pm1.5$ & $2.22\pm0.02$ & & & & &
  4.2 & 1.35[165]\\ 
PL                   &  15-200 & &
  $6.0\pm1.2$ & $2.02\pm 0.07$ & & & & &
  2.2 & 0.76[15]\\
BB+PL & 0.1-200 & $6.0\pm0.2$ &
  $5.2\pm0.4$ & $1.96\pm0.04$ & & & & 
  $0.50\pm0.02$ &
  4.6 & 0.97[163] \\
{\em comptt} & 0.1-200 & $3.5\pm 0.3$ & & &  
  $0.36\pm0.02$ & $107\pm 80$  &
  $0.6\pm0.5$ & &
  4.1 & 1.12[163]\\
BB+{\em comptt} & 0.1-200 & $3.3\pm 0.4$ & & & 
  $0.6\pm0.4$ & $77\pm93$  &
  $1.2\pm1.5$ & $0.43\pm0.09$ &
  4.3 & 0.99[161]

\enddata
\tablenotetext{a}{Energy range in keV}
\tablenotetext{b}{Value of Wisconsin absorption parameter in units of 
 $10^{21}$~cm$^{-2}$ } 
\tablenotetext{c}{Power law (PL) normalization at 1~keV, in units of $10^{-2}$}  
\tablenotetext{d}{Temperature of the comptonized soft seed photons, in keV}
\tablenotetext{e}{Plasma temperature, in keV}
\tablenotetext{f}{Plasma optical depth for disk geometry}
\tablenotetext{g}{Temperature of the blackbody (BB) component, in keV}   
\tablenotetext{h}{In units of $10^{-10}$ erg~cm$^{-2}$~s$^{-1}$}
\tablenotetext{i}{Number of d.o.f. is given in parentheses} 
\tablecomments{Errors are single parameter l$\sigma$ errors.}
\end{deluxetable}
    
\end{center}
\clearpage

                             
\begin{figure}
\plotone{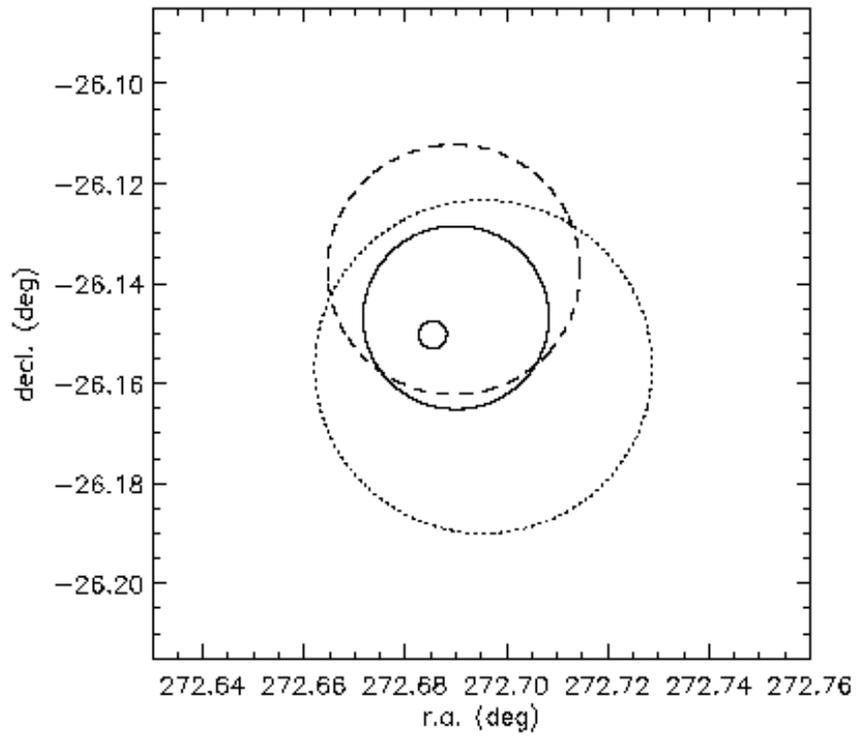}
\caption{
    Shown are the different source error circles (99\% confidence) as 
    derived from the detection of the burst with the {\em BeppoSAX}~WFC 
    (large circle
    thick line), of the persistent emission by the {\em BeppoSAX}~WFC 
    (dotted line), by the 
    NFI (broken line) and {\em ROSAT}~HRI (small circle thick line). 
   \label{fig1}}
\end{figure}

\clearpage

\begin{figure}
\plotone{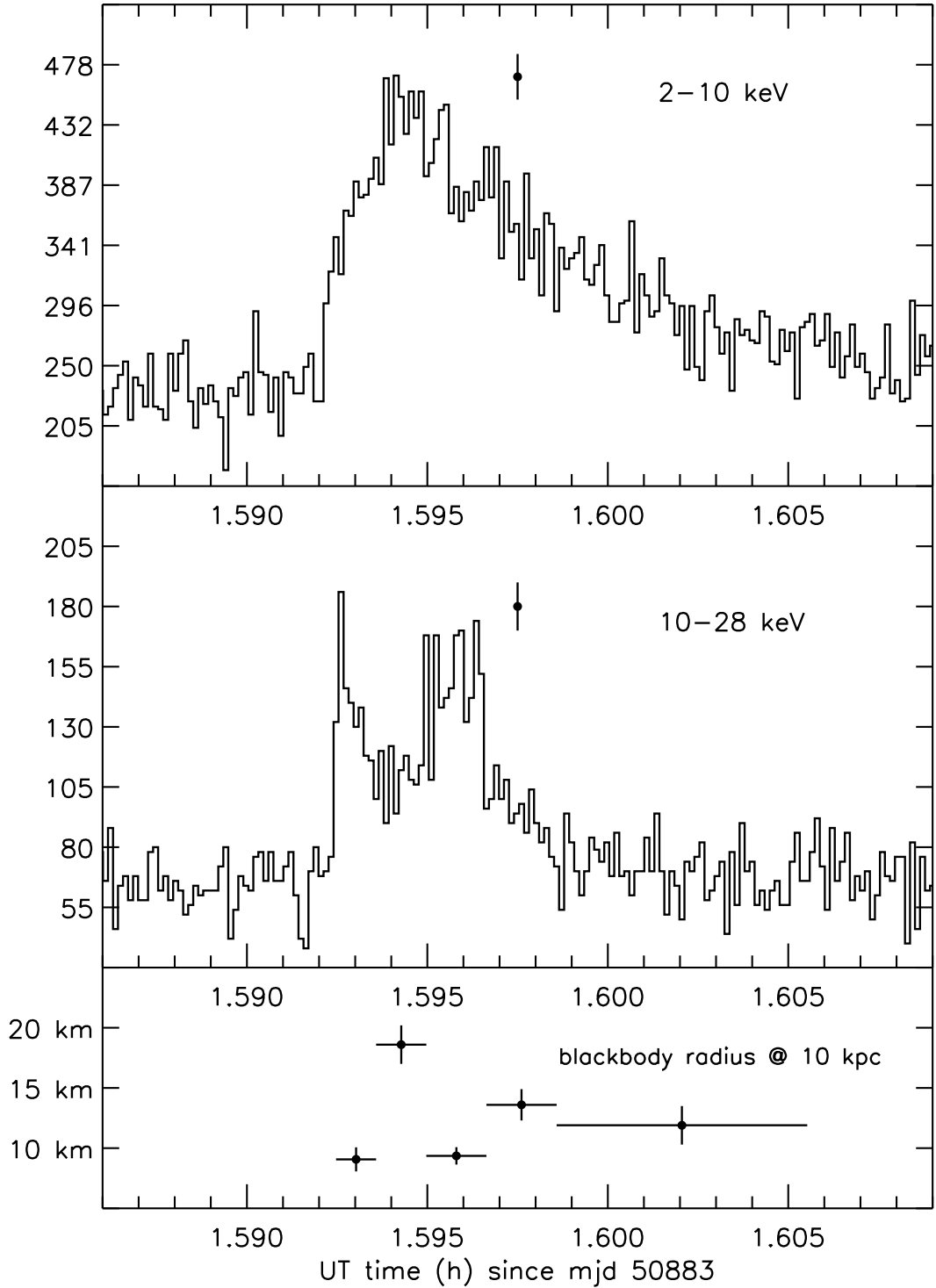}
\caption{
    Time profile in the low (2-10 keV, top panel) and high 
    (10-28 keV, middle panel) energy range 
    of the SAX J1810.8-2609 1998, March 11.06634 burst. The time 
    evolution of 
    the blackbody radius, computed for a distance of 10 kpc, is also shown.
    The actual blackbody radius is a linear function of source distance.       
   \label{fig2}}
\end{figure}

\clearpage

\begin{figure}
\plotone{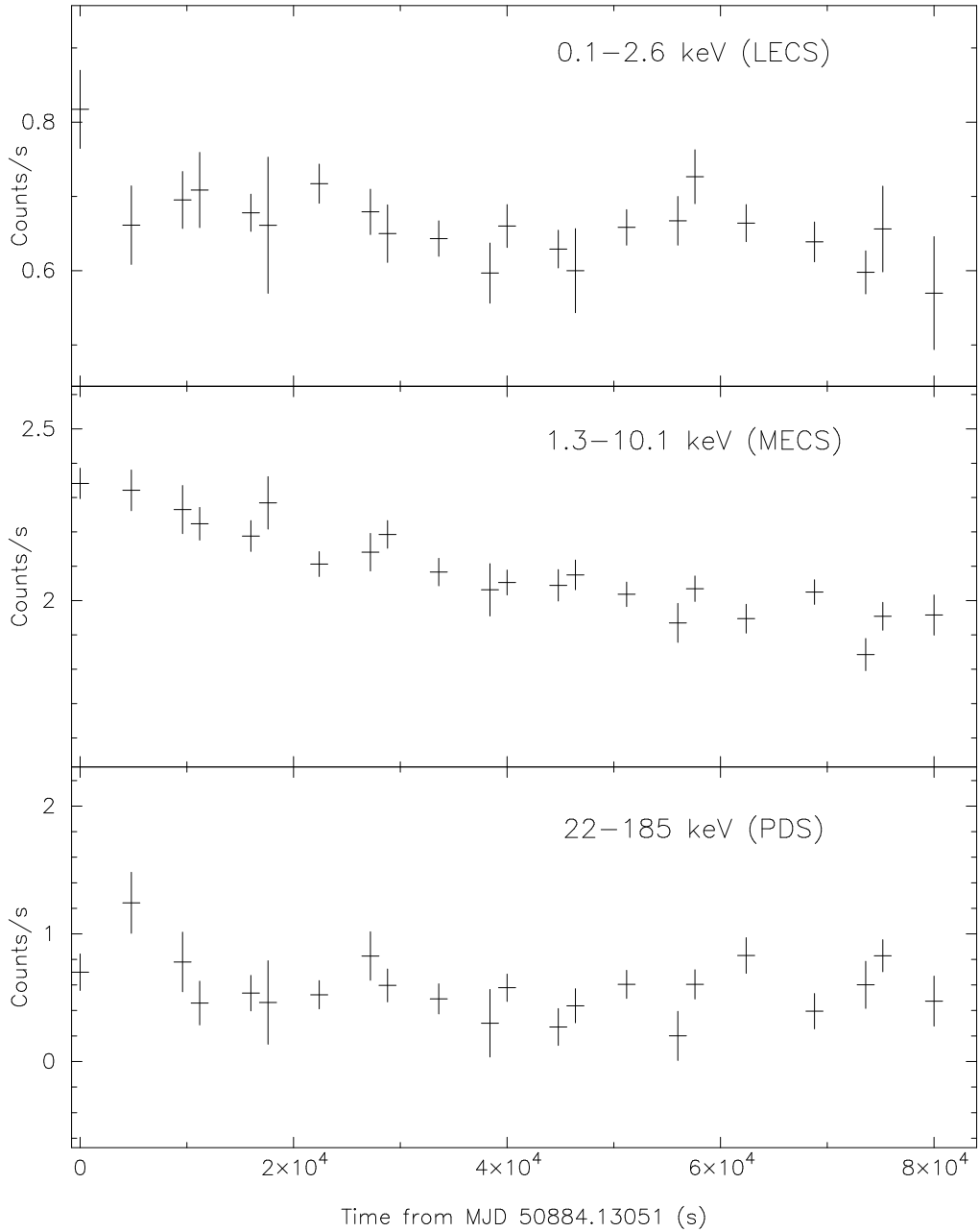}
\caption{
Time evolution of the X-Ray emission of SAX J1810.8-2609 in three 
different energy bands from the LECS (top), MECS (centre) and PDS (bottom).
A weak decline at low energies (E~$\leq10$~keV) 
is visible, while no variation is apparent 
in the hard band 
(22-185 keV). 
\label{fig3}}
\end{figure}

\clearpage

\begin{figure}
\plotone{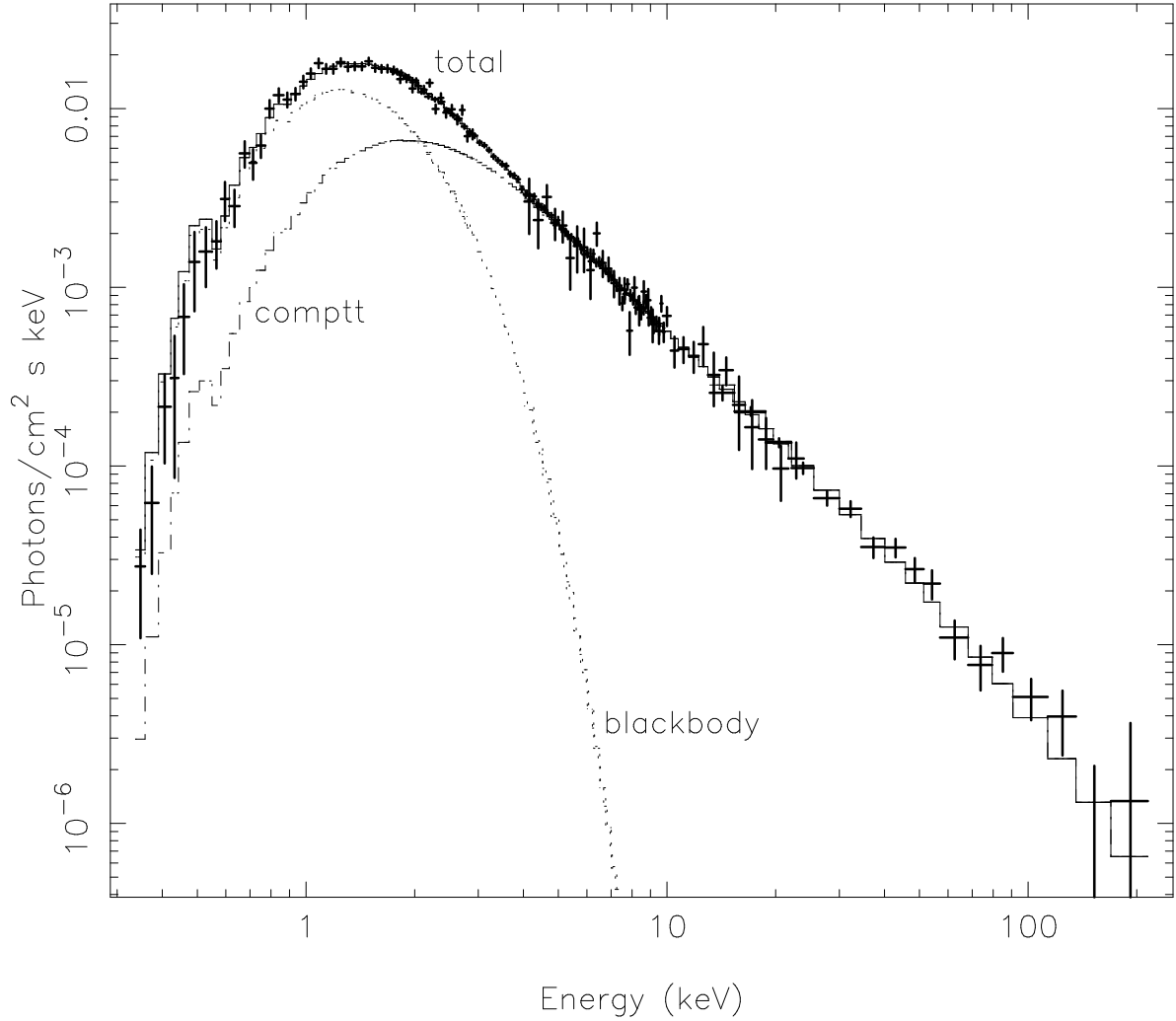}
\caption{
The unfolded broadband spectrum of SAX J1810.8-2609 as measured by the
{\em BeppoSAX} NFIs on March 12-13, 1998 during a 95 ksec exposure.
the best model fit to a single blackbody (dotted curve) plus thermal
comptonisation (dot-dashed curve) emission spectrum is shown.       
\label{fig4}}
\end{figure}


\clearpage

\begin{thebibliography}{}


\bibitem[Barret et al. 1999]{Bar99}
         Barret, D., et al. 1999, \apj, in press
\bibitem[Bazzano et al. 1997]{Baz97}
	Bazzano, A., et al. 1997, AIP Conf. Proc. N. 410, 729
\bibitem[Boella et al. 1997]{Boe97}
        Boella, G., et al. 1997, \aap 122, 327 
\bibitem[Christian \&\ Swank 1997]{Chr97}
         Christian, D.J. \&\ Swank, J.H. 1997, \apjs, 109, 177
\bibitem[Cocchi, et al. 1998]{Coc98}
         Cocchi, M., et al. 1998, Nucl. Phys. B (Proc. Suppl.) 69/1-3, 232
\bibitem[Cocchi et al. 1999]{Coc99}
         Cocchi, M., et al. 1999, "Proceedings of The Extreme Universe, 
         3nd Integral Workshop" 14-18 Sept. 1998, Taormina, Italy, in press
\bibitem[Dickey \&\ Lockman 1990]{Dic90}
         Dickey, J.M., and Lockman, F.J. 1990, \araa, 28, 215
\bibitem[Frontera et al. 1997]{Fro97}
	 Frontera, F., et al. 1997, \aap, 122, 327  
\bibitem[Ebisuzaki (1987)]{Ebi87}
         Ebisuzaki, E. 1987, \pasj, 39, 287
\bibitem[Greiner et al. 1998]{Gre98}
         Greiner, J., Castro-Tirado, A.J., \& Boller, T. 1998, IAU circ. 6985
\bibitem[Greiner et al. 1999]{Gre99}
         Greiner, J., et al. 1999, \mnras, 308, L17 
\bibitem[Guainazzi et al. 1998]{Gua98}
	 Guainazzi, M., et al. 1998, \aap, 225, 802
\bibitem[Heise et al. 1999]{Hei99}
         Heise, J., et al. 1999, in Proc. 3rd INTEGRAL Workshop, 
         {\em "The Extreme Universe"}, in press 
\bibitem[in~'t Zand et al.1999]{Zan99}
         in~'t Zand, J.J.M et al., 1999, \aap, 347, 891 
\bibitem[Jager et al. 1997]{Jag97}
         Jager, R., et al. 1997, \aap, 125, 557
\bibitem[Lewin, van Paradijs, \& Taam 1993]{Lew93} 
         Lewin, W.H.G., van Paradijs, J., \& Taam, R.E. 1993, Space Sci. Rev., 62, 223
\bibitem[Lewin et al. 1995]{Lew95}
 	Lewin, W.H.G., van Paradijs, J., \& Taam, R.E. 1995, in {\em "X-ray Binaries"},
        ed. W. Lewin, J. van Paradijs, \& E. van den Heuvel,
        Cambridge University Press, Cambridge, p. 175
\bibitem[Manzo et al. 1997]{Man97}
         Manzo, G., et al. 1997, \aap, 122, 341
\bibitem[Mitsuda et al. 1984]{Mit84}
         Mitsuda, K., et al. 1984, \pasj, 36, 741
\bibitem[Parmar et al. 1997]{Par97}
         Parmar, A.N., et al. 1997, \aap, 122, 309
\bibitem[Piraino et al. 1999]{Pir99}
         Piraino, S., et al. 1999, \aap, 349, L77
\bibitem[Revnivtsev et al. 1999]{Rev99}
         Revnivtsev, M., et al., 
         in Proc. 3rd INTEGRAL Workshop, {\em "The
         Extreme Universe"},          in press                              
\bibitem[Schulz 1999]{Sch99}
         Schulz, N.S., 1999, \apj, 511, 304
\bibitem[Sunyaev \&\ Titarchuk, 1980]{Sun80}
        Sunyaev, R.A., \&\ Titarchuk, L.G. 1980, \aap, 86, 121
\bibitem[Tavani \&\ Barret, 1997]{Tav97}
        Tavani, M. \&\ Barret, D. 1997, AIP Conf. Proc. N. 410, 75
\bibitem[Titarchuck 1994]{Tit94}
	 Titarchuck L., \aap, 434, 313
\bibitem[Ubertini et al. 1998a]{Ube8a}
         Ubertini P., et al., 1998a, IAU circ. 6838
\bibitem[Ubertini et al. 1998b]{Ube8b}
	 Ubertini P., et al., 1998b, IAU circ. 6843
\bibitem[Ubertini et al. 1999]{Ube99}
	 Ubertini P., et al. 1999, 
         in Proc. 3rd INTEGRAL Workshop, {\em "The
         Extreme Universe"}, in press 

\end{thebibliography}
\end{document}